\documentclass[preprint,floatfix]{revtex4-1}
\usepackage{lineno}

\usepackage{graphicx}
\usepackage{epsfig}
\usepackage{enumerate}
\usepackage{amssymb}
\usepackage{amsmath}
\usepackage{amsfonts}
\usepackage{dcolumn}
\usepackage{bm}
\usepackage{color}
\usepackage[colorlinks,linkcolor=blue,citecolor=blue,urlcolor=blue]{hyperref}
\usepackage[mode=text]{siunitx}
\usepackage{mathbbol}
\usepackage{soul}

\newcommand{\be}{\begin{equation}}
\newcommand{\ee}{\end{equation}}
\newcommand{\bea}{\begin{eqnarray}}
\newcommand{\eea}{\end{eqnarray}}

\def\d{\delta}
\def\D{\Delta}
\def\e{\epsilon}
\def\ve{\varepsilon}

\def\l{\lambda}

\def\n{\nu}

\def\p{\pi}

\def\r{\rho}

\def\t{\tau}

\def\F{\Phi}

\def\w{\omega}

\def\Q{\Psi}

\def\bld{{\mathbf d}}

\def\blk{{\mathbf k}}

\def\blq{{\mathbf q}}

\def\blU{{\mathbf U}}

\def\callO{\mbox{$\mathcal{O}$}}

\def\ua{\uparrow}
\def\da{\downarrow}

\def\bra{\langle}
\def\ket{\rangle}

\def\Tr{{\rm Tr}}

\def\1op{\hat{\mathbbm{1}}}

\def\AA{\mathring{\mathrm{A}}}

\begin{document}


\begin{center}
\textbf{\large{Theory of coherent phonons coupled to excitons}}
\vspace{0.5cm}

\vspace{0.5cm}

Enrico Perfetto$^{1,2}$, Kai Wu$^{1}$ and Gianluca Stefanucci$^{1,2}$

\vspace{0.5cm}

\textit{\footnotesize{$^1$Dipartimento di Fisica, Universit{\`a} di
Roma Tor Vergata, Via della Ricerca Scientifica 1,
00133 Rome, Italy\\
$^2$INFN, Sezione di Roma Tor Vergata, Via della Ricerca Scientifica
1, 00133 Rome, Italy}}\\
\vspace{0.5cm}

* \textit{Corresponding Author:}
Enrico Perfetto

\textit{Email address:} enrico.perfetto@roma2.infn.it

\vspace{1cm}

\end{center}

\textbf{
The interaction of excitons with lattice vibrations underlies 
the scattering from bright to dark excitons as well as the coherent 
modulation of the exciton energy. Unlike the former mechanism, which 
involves phonons with finite momentum,  the latter can be exclusively 
attributed to {\it coherent phonons} with zero momentum. 
We here lay down the microscopic theory of coherent phonons 
interacting with  resonantly pumped bright
excitons and provide the explicit expression of the corresponding 
coupling. The coupling notably resembles the exciton-phonon 
one, but with a crucial distinction: it contains the 
bare electron-phonon matrix elements rather than the screened ones.   
Our theory predicts that the exciton energy features a 
polaronic-like red-shift and 
monochromatic oscillations or beatings, 
depending on the number of coupled optical modes.
Both the red-shift and the
amplitude of the oscillations are 
proportional to the excitation density and to the 
square of the exciton-coherent-phonon coupling. We 
validate our analytical findings through comparisons with 
numerical  simulations of time-resolved optical absorbance 
in resonantly pumped MoS$_{2}$ monolayers.
}

\newpage

\section{Introduction}

Excitons, bound states formed by one electron and one hole, dominate the photophysics 
of a wide class of novel functional materials including transition metal 
dichalcogenides~\cite{Qiu2013,Chernikov2014,He2014,WangChernikov2018}, 
perovskites~\cite{PhysRevApplied.2.034007,perov2,perov3}, 
Xenes~\cite{xen0,xen1,xen3} and related van der Waals 
heterostructures~\cite{vdW1,vdW2,vdW3}.   
The microscopic understanding of the competing processes governing the exciton dynamics
is therefore of crucial importance for technological applications in 
optoelectronics~\cite{Wang2012,Mak2016,mullermalic}, 
photovoltaics~\cite{photo0,photo1,photo2} and 
photocatalysis~\cite{photoc1,photoc2,photoc3}.

The non-equilibrium properties of excitons are largely influenced by 
the electron-phonon  ({\it e-ph}) coupling $g$~\cite{giustino}, that  has been shown to 
determine excitonic relaxation 
lifetimes~\cite{lifetime1,lifetime2,lifetime3}, optical 
linewidths~\cite{optical1,Qiu2013}, energy 
renormalizations~\cite{renorm1}, dephasing 
rates~\cite{decoh1,decoh2,decoh3}, and diffusion dynamics~\cite{diff1,diff2}.
The theoretical description of the exciton dynamics takes advantage
from a very useful quantity  called exciton-phonon (X{\it-ph}) 
coupling~\cite{Toyozawa}.
Its expression  has been independently derived by a 
number of authors~\cite{Toyozawa,exph3,exph2,exph1,exph4} and 
it involves the projection of the {\it screened} 
{\it e-ph} coupling $\mathbb{g}\equiv =\ve^{-1}g$ (with $\ve$ the 
dielectric constant)
between two excitonic states. Accurate calculations of the X{\it-ph}
coupling have been performed and
excellent agreement between theory and experiments has been reported
for phonon-assisted photoluminescence 
spectra~\cite{exph3,PhysRevMaterials.7.024006},
exciton linewidths~\cite{PhysRevB.98.045143,exphlife}, valley 
depolarization time~\cite{PhysRevResearch.4.043203}, and polaronic redshifts
 in absorption spectra~\cite{PhysRevLett.119.187402}.

The X{\it-ph} coupling does not exhaust the possible interactions 
between excitons and lattice vibrations. Transient 
 optical spectroscopies
have recently
revealed periodic modulations of the 
excitonic 
resonances~\cite{trovatello,cohexp2,cohexp3,cohexp4,cohexp5}, 
pointing to a 
strong coupling between excitons 
and {\it coherent phonons} (X{\it-cph}), i.e. waves of atomic vibrations extending over a macroscopic spatial range.
In spite of the great interest in the topic, a microscopic theory of 
the X{\it-cph} interaction is still lacking.

In this work we show that the X{\it-cph} coupling
$G^{\l}_{\n}$ between  the lowest bright exciton $\l$  generated after 
resonant pumping
and a coherent phonon mode $\n$ is 
\begin{align}
G^{\l}_{\n}=
\sum_{\blk vcc'}g^{\nu}_{ c c'}(\blk) A^{\l*}_{\blk vc} 
A^{\l}_{\blk vc'}
-\sum_{\blk cvv'}g^{\nu}_{  v v'}(\blk) A^{\l*}_{\blk v'c} 
A^{\l}_{\blk vc},
\label{ecp}
\end{align}
where $A^{\l}_{\blk vc}$ is the exciton wavefunction,
$g^{\nu}_{ij} (\blk)=g^{\nu}_{ij} (\blk,\blq=0)$ is the {\em bare} {\it e-ph}
coupling for an electron with momentum $\blk$ to be scattered from 
band $i$ to band $j$ by the phonon mode $\n$ (momentum transfer $\blq=0$), and the sum runs over all 
valence ($v$) and conduction ($c$) bands. We further show that for 
resonant pumping the X{\it-cph} coupling is responsible 
to change the exciton energy in time according to (up to a phase) 
\begin{align}
\d E^{\l}(\t)= n\sum_{\n}\frac{|G^{\l}_{\n}|^{2}}{\hbar 
\w_{\n}}\times\big[\,\frac{n_{\n}}{n}\cos 
\w_{\n}\t-1\big],
\label{exct}
\end{align}
where $\w_{\n}\equiv \w_{\n \blq=0}$ is the frequency of the $\n$-th optical 
mode at the $\Gamma$ point, $n$ is the excitation density (number of conduction 
electrons per unit cell) and $n_{\n}\leq n$ are ``effective'' 
densities depending on the duration $T_{\mathrm{P}}$ of the pump pulse. 
The inequality is saturated ($n_{\n}=n$)
only for $T_{\mathrm{P}}$ shorter than 
the optical periods $T_{\n}=2\p/\w_{\n}$ (displacive 
excitations~\cite{PhysRevB.45.768,PhysRevLett.73.3243,ishioka2009coherent}).

The interaction with coherent phonons does not involve a scattering 
between different excitons, its main effect being 
a polaronic-like red-shift $-n\sum_{\n}\frac{|G^{\l}_{\n}|^{2}}{\hbar 
\w_{\n}}$ and 
monochromatic oscillations or beatings 
(depending on the number of coupled optical modes) of the exciton 
energy. The polaronic-like red-shift adds up to the one due to 
(incoherent) phonons~\cite{exphlife}, which involves the square of 
the X-{\it ph} coupling and it is proportional to the 
number of phonons rather than to the excitation density $n$.
Notably the X{\it-cph} coupling in Eq.~(\ref{ecp}) 
is formally identical to the X{\it-ph} 
coupling evaluated at the $\Gamma$ point for the diagonal scattering 
$\l\to\l$~\cite{Toyozawa,exph3,exph2,exph1,exph4}, with 
the crucial difference that the bare $g$ instead of 
the screened $\mathbb{g}$ appears in the former.
The difference arises from the diagrammatic origin
of these couplings, the X{\it-cph} one emerging from 
the Ehrenfest self-energy~\cite{stefanucci_in-and-out_2023}, see below.
From Eq.~(\ref{exct}) we infer that the amplitude of the coherent oscillations
in a time-resolved   optical spectrum is proportional 
to the excitation density and the square modulus of the X{\it-cph} 
coupling. This prediction is confirmed through  
real-time (RT) simulations of transient absorption 
in  resonantly pumped MoS$_{2}$ monolayer.
Our results provide a readily applicable formula 
for a quantitative understanding of the coherent dynamics 
in excitonic materials.

\section{Results}

{\bf Coupling between excitons and coherent phonons}

We consider a semiconductor hosting bound excitons,
and  denote by $\l$   the lowest  optically bright 
exciton. The exciton state is described by the 
coherent superposition of electron-hole pairs 
\be
| \l \ket = \sum_{\blk vc}A^{\l}_{\blk vc} \hat{d}^{\dag}_{\blk 
c}\hat{d}_{\blk v} 
|\Phi_{0}\ket ,
\label{excwf}
\ee
where the operator  $\hat{d}^{(\dag)}_{ \blk i}$ 
annihilates (creates)
an electron with momentum $\blk$ in band $i=\{v,c\}$, 
and $|\Phi_{0}\ket$ denotes the 
ground-state (filled valence bands  and empty
conduction bands). The exciton wavefunction
$A^{\l}_{\blk vc}$ is normalized to unity, $\sum_{\blk 
vc}|A^{\l}_{\blk vc}|^{2}=1$, and is obtained by solving the 
Bethe-Salpeter equation (BSE) $H A^{\l}=E^{\l}A^{\l}$ where
\be
H_{\blk vc,\blk' v'c'}=(\e_{\blk c}-\e_{\blk 
v})\d_{\blk \blk'}\d_{cc'}\d_{vv'} + K_{\blk vc,\blk' v'c'}.
\label{bseh}
\ee
Here $\e_{\blk v}$ and $\e_{\blk c}$ are valence and conduction band 
dispersions and $K$ is the BSE kernel in the Hartree plus 
statically screened exchange (HSEX) approximation. 

For weak resonant pumping with the exciton energy $E^{\l}$ the 
many-body quantum state at time $t$ is well described by 
$|\Q(t)\ket=|\F_{0}\ket+\sqrt{N(t)}\,e^{iE^{\l}t}|\l\ket$, where 
$N(t)$ is the total number of conduction electrons at time $t$. 
The weak pumping assumption implies that the excitation density 
$n(t)=N(t)/N_{k}\ll 1$, $N_{k}$ being the number of $\blk$-points
in the first Brillouin zone.
The change in the one-particle density matrix to lowest 
order in $n$ is then given by
\begin{subequations}
\begin{align}
\d \r_ {\blk c c'}(t) &= \bra  \Psi(t) | 
 \hat{d}^{\dag}_{\blk c'}\hat{d}_{\blk c} |\Psi(t) \ket = 
 N(t) \sum_{v} A^{\l *}_{\blk vc'} A^{\l}_{\blk vc} , 
\label{drhocc}
 \\
 \d \r_ {\blk v v'}(t) &= -\bra \Psi(t) | 
\hat{d}_{\blk v} \hat{d}^{\dag}_{\blk v'} |\Psi(t) \ket
 =-N(t) \sum_{c} A^{\l *}_{\blk vc} A^{\l}_{\blk v'c} .
\label{drhovv}
\end{align}
\label{drho}
\end{subequations}

\noindent
Notice that for times subsequent to the end of the pump, i.e., 
$t>T_{\mathrm{P}}$, $N(t)=N$ is 
independent of time~\cite{PSMS.2019,PhysRevB.103.245103}, and so are 
the density matrix elements for each $\blk$.
This peculiarity arises from pumping resonantly with the lowest 
excitonic state.

The bare {\it e-ph} coupling governs the dynamics of coherent 
phonons. Let 
$x_{\n}$ be the displacement of the $\n$-th 
optical mode. To lowest order in $n$ and for times 
$t>T_{\rm P}$
it satisfies the equation of motion 
(EOM)~\cite{stefanucci_semiconductor_2024}
\begin{align}
M\ddot{x}_{\n}(t) &=-M\w^{2}_{\n} x_{\n}(t)
-\frac{1}{x_{0\n}N_{k}} \sum_{\blk i j} g^{\nu}_{i j}(\blk) \d 
\r_{\blk j i}(t),
\label{eqx}
\end{align}

\noindent
with $M$ the mass of the unit cell and $x_{0\n}=\sqrt{\hbar/(M\w_{\n})}$.
For resonant pumping we can solve Eqs.~(\ref{eqx}) 
with $\d\r$ from Eqs.~(\ref{drho}). Indeed the phononic feedback on 
the electronic density matrix yields a correction of order 
$\callO(n^{2} g^{2})$
that can be  neglected. This allows us to rewrite the  
EOM as
\begin{align}
\ddot{x}_{\n}(t)=-\w^{2}_{\n} x_{\n}(t)-n(t) 
\frac{
G^{\l}_{\n}}{\hbar}\w_{\n}\,x_{0\n},
\label{ddotx}
\end{align}
where the X{\it-cph} coupling $G^{\l}_{\n}$ has been defined in 
Eq.~(\ref{ecp}). The appearance of the bare {\it e-ph} coupling $g$ 
in $G^{\l}_{\n}$ implies that the interaction between excitons and coherent 
phonons can be substantially larger  than the interaction between  
excitons and phonons~\cite{PhysRevB.107.024305}.

Equation (\ref{ddotx}) admits an analytic solution for any 
time-dependent excitation density $n(t)$. Let us discuss some 
physically relevant limiting cases. For displacive 
excitations~\cite{PhysRevB.45.768,PhysRevLett.73.3243,ishioka2009coherent}, i.e., if the duration $T_{\mathrm{P}}$
of the pump pulse 
is much shorter than the phononic period $T_{\n}=2\pi/\w_{\n}$, then we can 
set $n(t)=n$ and solve Eq.~(\ref{ddotx}) with initial 
conditions $x_{\n}(T_{\mathrm{P}})=\dot{x}_{\n}(T_{\mathrm{P}})=0$. The solution is 
\be
x_{\n}(t)=x_{0\nu }\frac{n G^{\l}_{\n}}{\hbar \w_{\n}} [\cos 
\w_{\n}(t-T_{\mathrm{P}})-1].
\label{ut}
\ee
Equation~(\ref{ut}) shows that fast resonant pumping produces coherent 
oscillations of the nuclear displacements 
with amplitude $a_{\n}=|\mathsf{x}_{\n}|\equiv\frac{x_{0\nu }n|G^{\l}_{\n}|}{\hbar 
\w_{\n}}$.
In addition, the oscillations occur around an average displaced position 
$\mathsf{x}_{\nu}=-\mathrm{sign}(G^{\l}_{\n}) |\mathsf{x}_{\nu}|$, that can be 
positive or negative, depending on the 
sign of the X{\it-cph} coupling.
If, on the other hand, the pump duration is  comparable or larger than the phonon 
period then the average displaced position does not change 
while the amplitude of the oscillations reduces:
$a_{\n}=(n_{\n}/n)|\mathsf{x}_{\n}|$ with $n_{\n}<n$.
The ``effective'' 
densities $n_{\n}=n_{\n}(T_{\rm P})$ are decreasing functions of 
$T_{\rm P}$ which approach $n$ for $T_{\rm P}\ll T_{\n}$ and zero 
for $T_{\rm P}\gg T_{\n}$. In fact, 
in the limit of very slow pumping  
the displacement $x_{\n}(t)\approx \mathsf{x}_{\nu}$ becomes 
independent of time, consistently
with the fact that the system  is
adiabatically driven toward a nonequilibrium steady-state.

{\bf Coherent modulation of the exciton energy}

We now address how coherent phonons modify the exciton 
energy $E^{\l}$ as measured in a transient optical
experiments~\cite{trovatello,cohexp2,cohexp3,cohexp4,cohexp5}. We are interested in probing the system at a time $\t$
subsequent the phonon-induced dephasing of the electronic 
polarization~\cite{decoh1,dephasing2,dephasing1} 
(typical time-scales are a few hundreds of femtoseconds). 
In this depolarized regime the inter-band 
elements of the density matrix vanish~\cite{perfetto_2023}, i.e., $\r_{\blk vc}=0$, and 
the system is in a nonequilibrium steady-state 
characterized by a density matrix  
$\r =\r_{\rm{eq}}+\d \r$, with $\r_{\rm{eq}}$ the ground state 
density matrix and $\d \r$ as in Eqs.~(\ref{drho}).
The quasi-particle Hamiltonian 
then reads 
$h^{\mathrm{qp}}_{\blk}(\t)=
h^{\mathrm{HSEX}}_{\blk}[\r]+h^{\mathrm{Eh}}_{\blk}(\t)$, 
where the first term is the  HSEX 
Hamiltonian (which is a functional of the time-independent
density matrix $\r$) and 
the second term is the {\it time-dependent} Ehrenfest potential 
given by
$h^{\mathrm{Eh}}_{\blk ij}(\t)=\sum_{\n} g^{\nu}_{  i j}(\blk) 
\frac{x_{\nu}(\t)}{x_{0\nu}}$.
The Ehrenfest potential arises from 
the photo-excited coherent motion of the nuclei and acts on the 
electrons with an intensity proportional to the bare {\it e-ph} 
couplings $g$~\cite{stefanucci_in-and-out_2023}.

For probe pulses shorter than the optical phonon periods the 
transient absorption spectrum can be obtained using the adiabatic 
approach   of Ref.~\cite{PSMS.2015}. The absorption 
energies at time $\t$ are the eigenvalues of the non-equilibrium 
BSE with Hamiltonian $H(\t)=H+\d H(\t)$ with 
$\d H(\t)=\d H^{\rm Eh}+\d K$.
The change $\d K $ of the BSE kernel $K=\d h^{\rm HSEX}/\d\r$ in 
Eq.~(\ref{bseh}) is due to
the pump-induced change of the density matrix~\cite{PSMS.2015}.
The term $\d H^{\rm Eh}$ does instead accounts for the renormalization of the 
quasi-particle energies caused by the  Ehrenfest 
potential. It is responsible for the temporal modulation
of the excitonic  energies and reads 
\be
\d H^{\rm Eh}_{\blk vc,\blk'v'c'}(\t)=\d_{\blk\blk'}\sum_{\nu}
\frac{x_{\n}(\tau)}{x_{0\n}}
\left[\d_{vv'}  g^{\nu}_{c c'}(\blk) -\d_{cc'}  g^{\nu}_{v' v}(\blk) 
\right].
\ee
 For small excitation densities $n$, both $\d H^{\rm Eh}$
and $\d K$  can be treated perturbatively, 
yielding the $\t$-dependent modification of
the $\l'$ excitonic energy
\bea
E^{\l'}(\t)&=&E^{\l'}+\sum_{\blk \blk'cc' vv'}A^{\l' *}_{\blk vc} [\d 
H_{\blk vc,\blk' v'c'}(\t)] A^{\l'}_{\blk' v'c'} \nonumber \\
&\equiv& E^{\l'}+\d E^{\l'}_{K}+\d 
E^{\l'}(\t).
\label{overallEl}
\eea 
The constant shift $\d E^{\l'}_{K}$ is due to $\d K$, and it accounts 
for the renormalization  caused by electronic 
correlation
effects like band-gap 
renormalization~\cite{Chernikov2015,Pogna.2016,Yao2017}, Pauli 
blocking~\cite{Steinhoff2014,Pogna.2016,smejkal2020time}
and excited-state screening~\cite{Steinhoff2014,Steinhoff2017,smejkal2020time}.
The explicit evaluation  of $\d E^{\l'}_{K}$ has been  the subject of 
several papers and is not  discussed further here.
The time-dependent shift $\d E^{\l'}(\t)$ origins from
$\d H^{\rm Eh}$.
Using the expression for $x_{\n}(\t)$ in Eq.~(\ref{ut}), derived  for 
$T_{\rm P}\ll T_{\n}$,
we obtain that the modulation of the $\l'$ exciton energy due to 
 resonant pumping with the lowest exciton  reads
\begin{align}
\d E^{\l'}(\t)= n\sum_{\n}\frac{G^{\l}_{\n} G^{\l'}_{\n}}{\hbar 
\w_{\n}}\times\big[\,\frac{n_{\n}}{n}\cos 
\w_{\n}\t-1\big],
\label{exctllp}
\end{align}
 that for $\l'=\l$ reduces to Eq.~(\ref{exct}).

We conclude that 
the coupling with coherent phonons in resonantly excited materials
is responsible for oscillations of the 
excitonic peak versus the impinging time $\t$ of the probe~\cite{trovatello,cohexp2,cohexp3,cohexp4,cohexp5}.
The oscillation periods are dictated by the phonon frequencies and, 
for the lowest-energy exciton, 
the amplitude is proportional to the square of the X{\it-cph} coupling.  
Our result also predicts a likelihood of observing 
beatings when more phonons
with different frequencies are strongly coupled to the exciton. This finding 
aligns with the outcomes of recent experiments in MoSe$_{2}$~\cite{edgeph}.

{\bf Transient absorption in monolayer MoS$_{2}$}

  \begin{figure}[tbp]
      \includegraphics[width=0.7\textwidth]{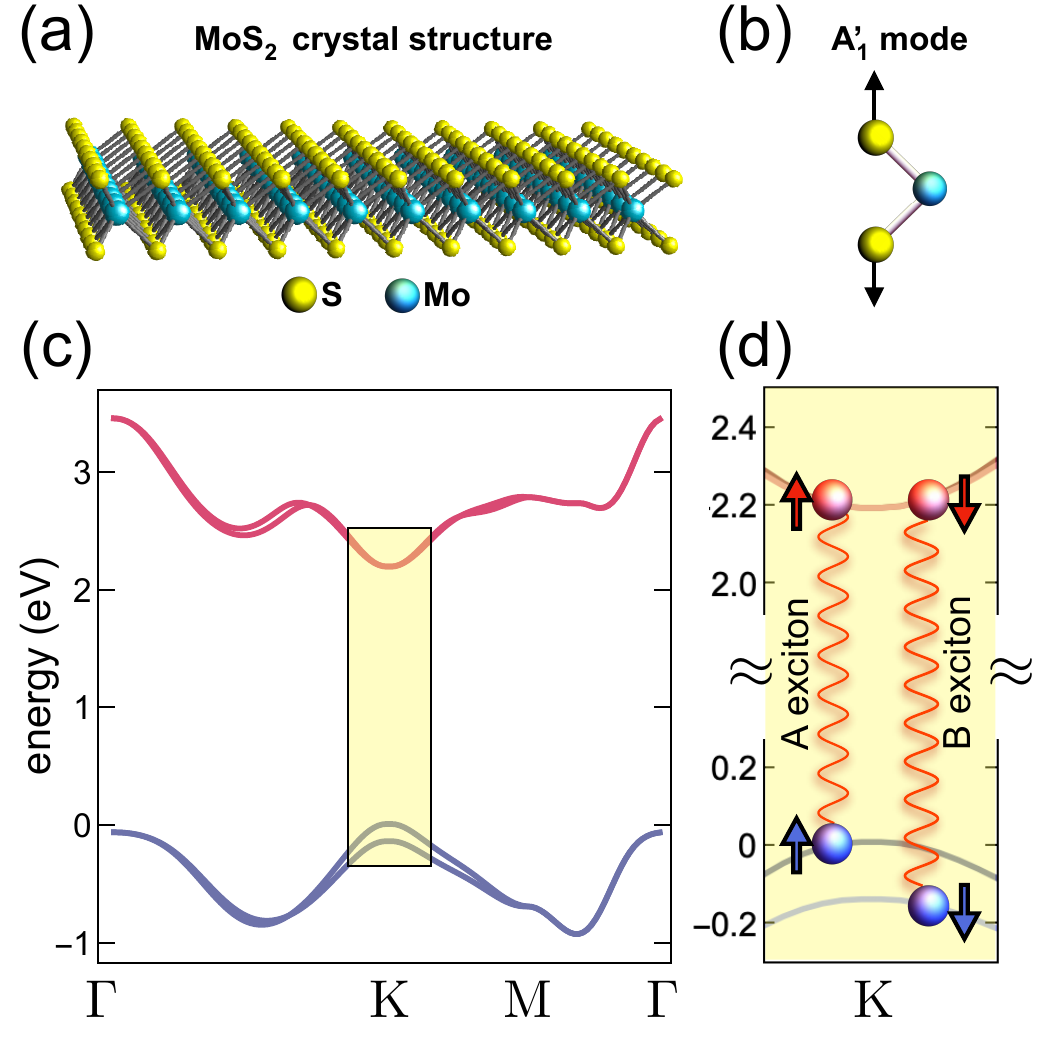}
   \caption{(a)~Crystal structure of a MoS$_{2}$ monolayer. 
   (b)~Geometrical representation of the $A_{1}'$ normal mode.
   (c)~Electronic band structure. The yellow box at 
   the $K$-point highlights the splitting due to spin-orbit 
   interaction. (d) Pictorial view of the bands involved in the 
   formation of A and B excitons.  
   }
   \label{fig1}
   \end{figure} 
We validate our analytical findings through 
numerical simulations of the transient absorbance in 
monolayer MoS$_{2}$, see Fig.~\ref{fig1}a.
We use a tight-binding description of the band structure~\cite{PhysRevB.88.085433}
and a Rytova-Keldysh potential for the 
screened interaction~\cite{PhysRevB.84.085406}, and  
consider only the out-of-plane $A'_{1}$ normal mode.
Recent experiments suggest that the photoexcited 
dynamics is dominated by such mode~\cite{trovatello}, see Fig.~\ref{fig1}b.
 \begin{figure}[tbp]
      \includegraphics[width=0.7\textwidth]{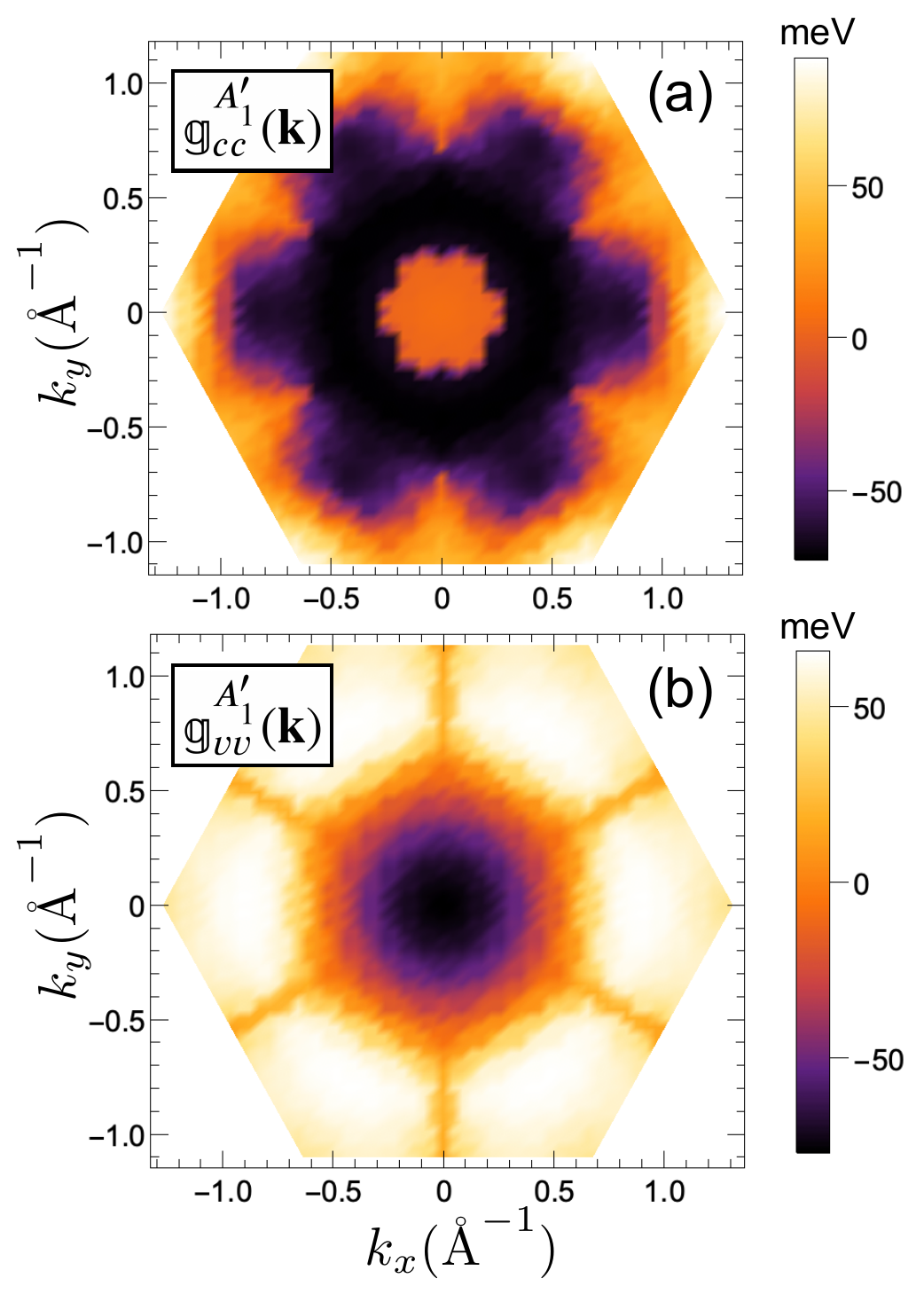}
   \caption{Color plot of the coupling between the $A_{1}'$ 
   optical phonon 
   and electrons in the conduction 
   (top) and valence (bottom) bands. 
   }
 \label{fig3}
\end{figure} 
For the bare {\it e-ph} coupling we use $g \approx \ve 
\mathbb{g}$ with dielectric constant $\ve=1$ --
screening at vanishing momentum transfer is negligible in 
two-dimensional 
materials~\cite{PhysRevB.84.085406,PhysRevB.92.245123}.
The screened {\it e-ph} couplings $\mathbb{g}^{A_{1}'}_{cc'}$ and 
$\mathbb{g}^{A_{1}'}_{vv'}$ have been evaluated with the {\tt Quantum 
Espresso} package~\cite{QuantumEspresso}, and are 
displayed in Fig.~\ref{fig3}.
We propagate in time the electronic 
density matrix $\r_{\blk}(t)$ in the HSEX+Ehrenfest approximation using the {\tt 
CHEERS} code~\cite{PS-cheers} 
(see Methods). 

The dynamics is initiated by a linearly polarized 
pump pulse of duration $T_{\rm P}=20$~fs (much shorter than the 
optical period $T_{\n=A_{1}'}\simeq 82$~fs)
and energy $\hbar \w_{\rm P}=1.9$~eV,
which is in resonace with the A exciton, see Fig.~\ref{fig1}c-d.
The pump pulse does therefore induce a displacive excitation and we 
expect $n_{\n=A_{1}'}\simeq n$. The excitation density for $t>T_{\rm P}$ 
is found to be $n=0.0013$, corresponding to a carrier density per 
unit area $n_{{\rm{x}}}=n/A_{\rm{MoS}_{2}}=1.5\times 10^{12}~\mathrm{cm}^{-2}$
($A_{\rm{MoS}_{2}}=8.8~ \AA^{2}$ being the area of the unit cell).
About 400~fs after the photoexcitation the  system enters  the 
depolarized regime with $\r_ {\blk v c}\approx 0$ and $\r_ {\blk c c'}$
and $\r_ {\blk v v'}$ independent of time. 
  \begin{figure}[tbp]
      \includegraphics[width=0.7\textwidth]{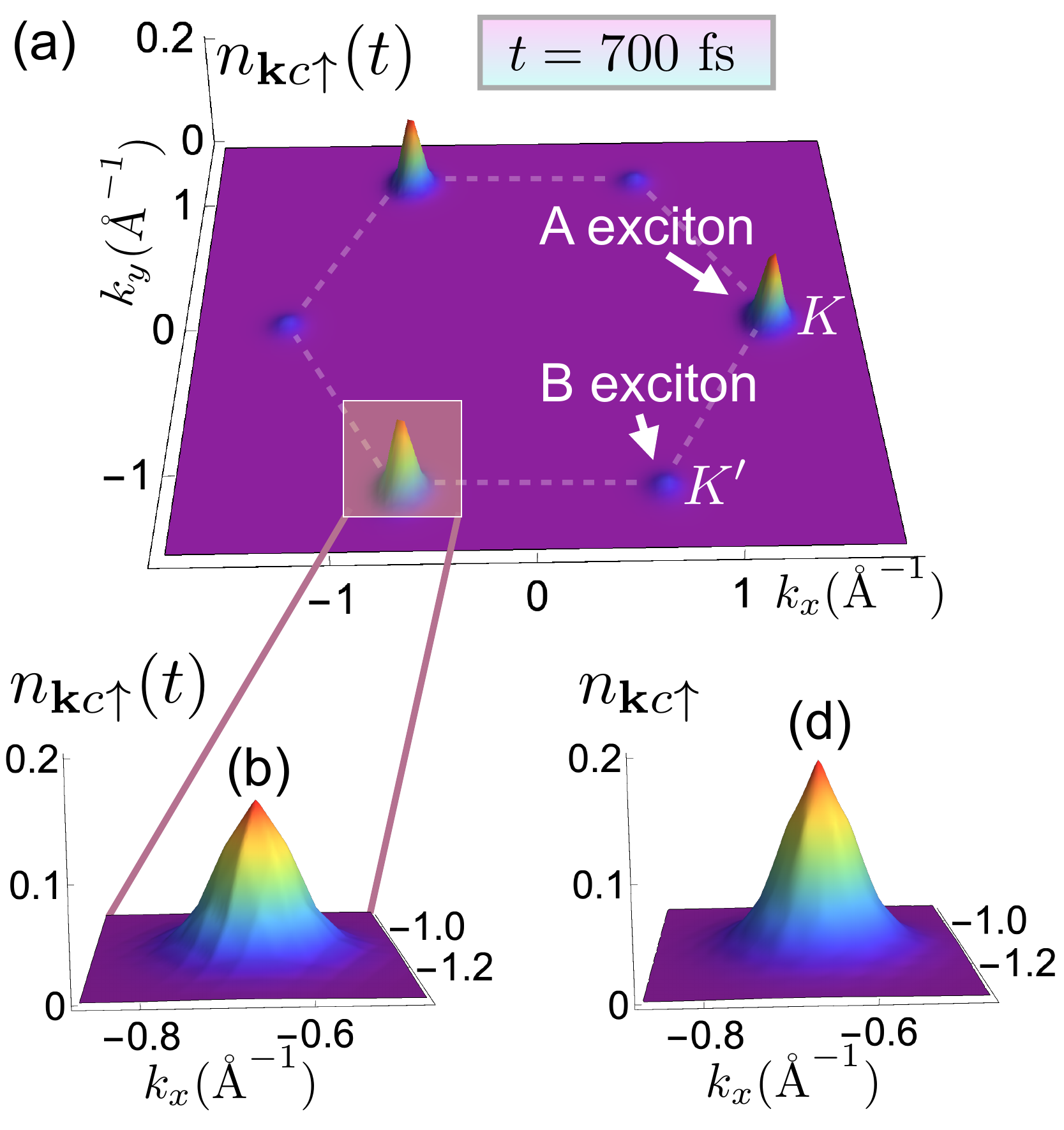}
   \caption{(a)~Plot of the density of spin-up electrons in the 
   conduction band after 700~fs (depolarized regime), showing a 
   predominance of the A exciton ($K$-point) over the B exciton 
   ($K'$point). (b)~Magnification of panel~(a) around the $K$-point. 
   (c)~Analytic density of spin-up electrons in the 
   conduction band around the $K$-point as obtained from 
   Eq.~(\ref{drhocc}).}
   \label{fig2}
   \end{figure} 
In Fig.~\ref{fig2}a 
we plot the occupations of spin-up electrons 
in the conduction band, i.e.,  
$n_{\blk c\ua}(t)=\r_{\blk c \ua c \ua}(t)$, at $t=700$~fs.
In the spin-up sector the A and B excitons are located at the 
$K$ and $K'$   valleys respectively, see again Fig.~\ref{fig1}c-d,  whereas in the 
spin-down sector this configuration reverses~\cite{WangChernikov2018}. 
In Fig.~\ref{fig2} we compare $n_{\blk c\ua}(t)$ (panel b) with 
the analytic value $n_{\blk c\ua} \equiv \d \r_ {\blk c \ua c \ua}$ 
from Eq.~(\ref{drhocc}) (panel c), which is proportional to the square 
modulus of the exciton wavefunction. 
The agreement is fairly good, thus confirming that resonant pumping 
generates a finite density of A excitons.  
The small discrepancy is due to the finite duration of the pump, which 
prevents achieving a perfect resonant condition. 
As a result, a minimal contamination arises from the generation of a 
small density of B excitons, specifically observed in Fig.~\ref{fig2}a 
at the $K'$ valley.

\begin{figure}[tbp]
      \includegraphics[width=0.6\textwidth]{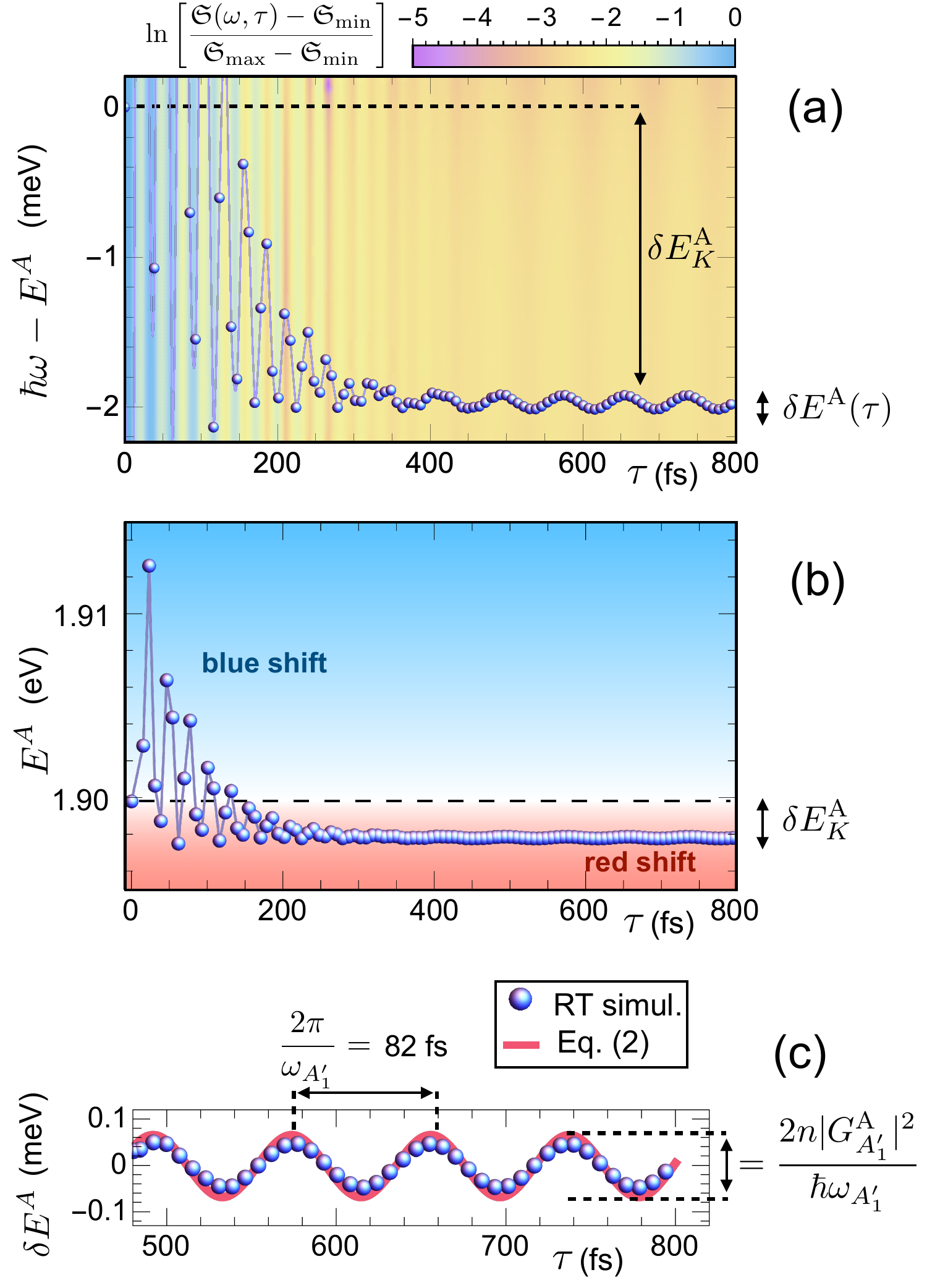}
   \caption{(a)~Color plot of the logarithmic transient absorption spectrum 
   $\ln\left[ \frac{\mathfrak{S}(\omega,\tau) -
   \mathfrak{S}_{\rm{min}}}{\mathfrak{S}_{\rm{max}}-\mathfrak{S}_{\rm{min}}} \right]$
   versus energy (vertical axis) and delay 
   (horizontal axis). Here $\mathfrak{S}_{\rm{min}}$ and 
   $\mathfrak{S}_{\rm{max}}$ denote the minimum and maximum values of 
   $\mathfrak{S}(\omega,\tau)$
   in the displayed region. Blue dots indicate the position of the maxima 
   of $\mathfrak{S}(\w,\t)$ for different delays. The electronic 
   red-shift $\d E^{\rm A}_{K}$ and the phonon-induced coherent modulation $\d 
   E^{\rm A}(\t)$ are also indicated. (b)~Temporal 
   evolution of the exciton energy $E^{{\rm A}}(\t)$ as a function of 
   the pump-probe delay $\t$. The reference equilibrium 
   value $E^{\rm A}(0)=1.9$~eV is shown as a dashed line, while blue 
   and red backgrounds are used visualize blue- and red-shift regions 
   respectively. (c)~Comparison of the position of the exciton 
   peak as obtained from the real-time (RT) simulations of panel~(a) 
   (blue dots) and the
   analytic formula in Eq.~(\ref{exct}) (red curve).
   }
   \label{fig5}
   \end{figure} 
   
To obtain the time-resolved absorption spectrum, 
we further imping 
the system with a broadband probe pulse 
at different delay times $\t$. We then calculate for 
all times $t$ the 
change in the density matrix 
$\D\r(t,\t)=\r_{\rm P+p}(t,\t)-\r_{\rm P}(t)$,  resulting 
from a simulation with pump and probe and a simulation with only the 
pump~\cite{PSMS.2015,PS.2015}.
The transient absorption spectrum  is calculated as~\cite{PSMS.2015,PS.2015}
\begin{align}
\mathfrak{S}(\w,\t)=-2\w\rm{Im}[\bf{e}(\w,\t) \cdot 
\bf{d}(\w,\t)],
\label{neqspectrum}
\end{align}
where $\bf{e}(\w,\t)$ 
is the Fourier transform of the probe field 
 and $\bf{d}(\w,\t)$ is the Fourier
transform of the probe-induced dipole moment ${\bf d}(t,\t)$.
The latter is related to the change of the density 
matrix via ${\bf d}(t,\t)=\sum_{\blk}\Tr[{\bf d}_{\blk} \D \r_{\blk} 
(t,\t) ]$, where ${\bf d}_{\blk}$ is the dipole transition matrix. 
Equation~(\ref{neqspectrum}) generalizes the equilibrium absorption 
spectrum to out-of-equilibrium conditions. 
In the absence of pump fields (equilibrium condition)  
we have ${\bf d}(\w,\t)= {\rm Tr}[{\bm 
\chi}(\w)\cdot {\bf e}(\w) ]$ independent of $\t$, with 
${\bm \chi}(\w)$ the 
dipole-dipole response function, 
and $\mathfrak{S}(\w,\t)$ reduces to equilibrium absorption 
spectrum~\cite{PhysRevB.84.245110,PSMS.2015}.

We have conducted simulations for several equidistant delay times 
$\t_{n}=n\D_{\t}$, 
encompassing approximately 100 distinct values
spanning a relatively long time-window ($\sim 1$~ps).
The dependence of the spectrum on $\t$ originates from 
the time-dependent HSEX+Ehrenfest 
potential: the specific timing at which the probe interacts 
with the system determines 
the value of  $h^{{\rm qp}}(\t)$, resulting 
in different frequencies for $\D\r_{\blk vc}(t,\t)$.

\begin{figure}[tbp]
      \includegraphics[width=0.7\textwidth]{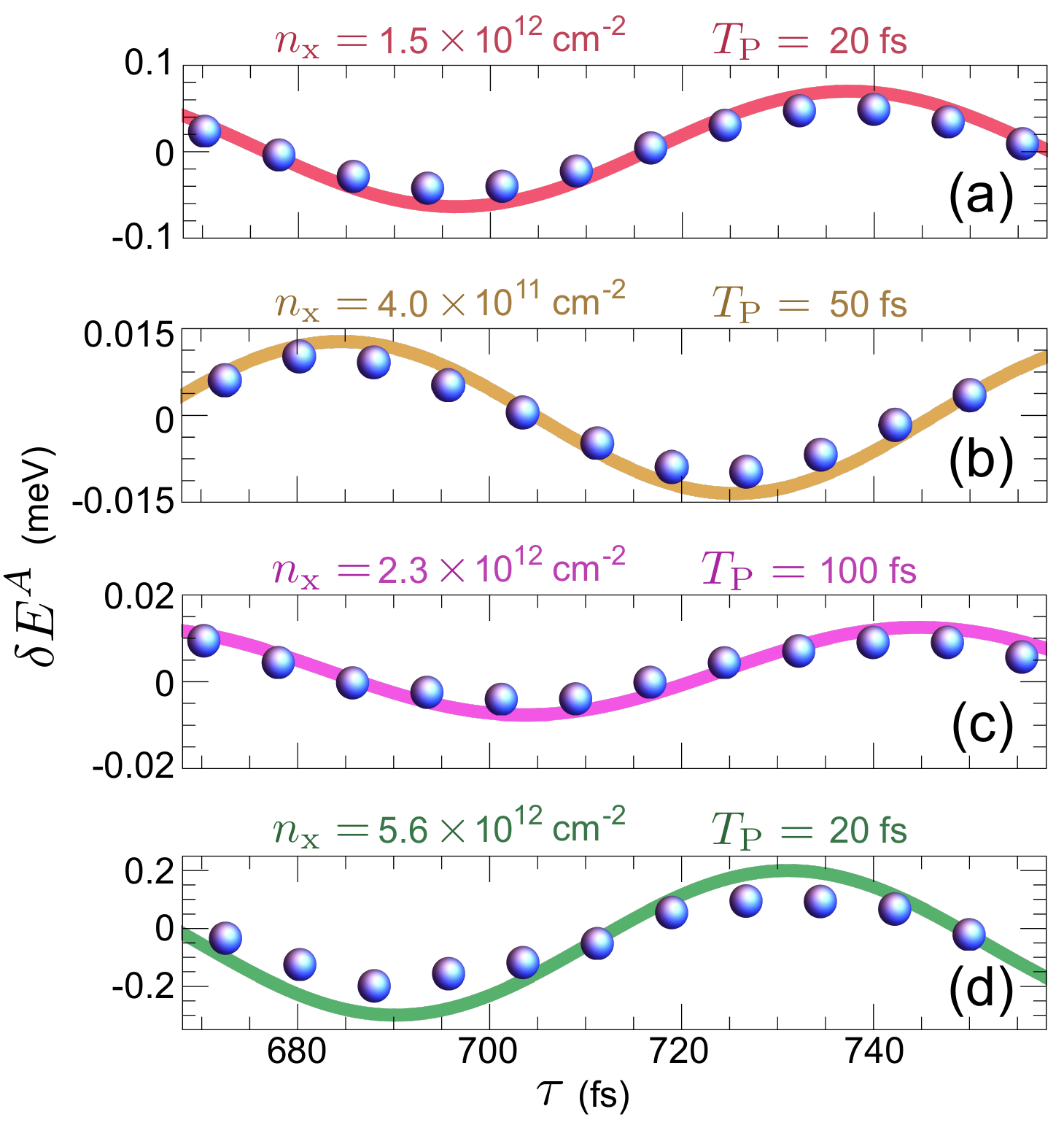}
   \caption{Comparison of the oscillations of the exciton 
   peak $\d E^{A}$ from RT simulations (blue dots) with the analytic 
   result of Eq.~(\ref{exct}) (solid curves) for different pump 
   dutations $T_{{\rm P}}$ and excitation 
   densities $n_{{\rm x}}$. 
   Panel a: $T_{{\rm P}}=20$~fs, $n_{{\rm 
   x}}=1.5\times 10^{12}~{\rm cm}^{-2}$ ($n_{A_{1}'}=n$);
   Panel b: $T_{{\rm P}}=50$~fs, $n_{{\rm 
   x}}=4.0\times 10^{11}~{\rm cm}^{-2}$ ($n_{A_{1}'}=0.71n$);
   Panel c: $T_{{\rm P}}=100$~fs, $n_{{\rm 
   x}}=2.3\times 10^{12}~{\rm cm}^{-2}$ ($n_{A_{1}'}=0.1n$);
   Panel d: $T_{{\rm P}}=20$~fs, $n_{{\rm 
   x}}=5.6\times 10^{12}~{\rm cm}^{-2}$ ($n_{A_{1}'}=n$);}
   \label{fig6}
   \end{figure}

In Fig.~\ref{fig5}a we display the logarthimic plot of $\mathfrak{S}(\w,\t)$ 
for energies 
$\hbar\w$ up to a few meV below the exciton energy $E^{\rm 
A}\simeq 1.9$~eV. The spectrum changes wildly until the  
pump-induced polarization is sizable (coherent regime), i.e., up to $\tau\simeq 
400$~fs. 
Both the position and the amplitude of the excitonic peak
oscillate with a period of $T^{\rm A}=2\p/ E^{\rm A}\sim 2.2$~fs.
In our simulations, however, the system is probed every $\D_{\t}=7.75$~fs, 
preventing us to resolve the faster  time-scale $T^{\rm A}$. 
Instead, an artificial period  
$T\sim 31$~fs [corresponding to the matching condition 
$T\sim pT^{\rm A}\sim q\D_{\t}$ with $p$ and $q$ prime numbers] is observed.

The position $E^{\rm{A}}(\t)$ of the A exciton energy, calculated as the 
maximum ${\rm max}_{\w}\mathfrak{S}(\w,\t)$, is shown in 
Fig.~\ref{fig5}b  (blue dots). The exciton is initially 
blue-shifted with respect to its equilibrium value. 
This blue-shift undergoes relaxation within about 200 fs, 
eventually transitioning into a stable red-shift $\d E_{K}^{\rm{A}}\approx 
2$~meV due to electronic mechanisms (change of the BSE 
kernel), see Eq.~(\ref{overallEl}). 
The decay time of this transition is governed by the polarization 
lifetime which, in turns, depends strongly on the excitation density
and it can vary from hundreds of femtoseconds
to picoseconds.
Our observations align closely with the work by Calati 
et al.~\cite{calati_ultrafast_2021}, where a similar 
blue-to-red transition has been reported in resonantly excited ${\rm 
WS}_{2}$.
Entering the depolarized (or incoherent) regime the 
wild oscillations gradually attenuate, and the 
intensity of the signal begins to oscillate at the phonon 
frequency $\w_{A_{1}'}$, see Fig.~\ref{fig5}a.
For $\tau> 400$~fs, the exciton position $\d E^{\rm{A}}(\t)$ features 
oscillations at the very same frequency 
 around the red-shifted value $ 
E^{\rm{A}}+\d E_{K}^{\rm{A}}$, in qualitative 
agreement with Eq.~(\ref{overallEl}) and Eq.~(\ref{exct}). 

Inspecting the amplitude of the oscillations, we also appreciate
an excellent quantitative agreement, see Fig.~\ref{fig5}c.
The red curve is the plot of Eq.~(\ref{exct}) with $n_{A_{1}'}=n$ 
(displacive excitation) and $n=0.0013$ the excitation density 
created by the pump.
The value of the {\it X-cph} coupling is 
$G^{\rm{\rm A}}_{A_{1}'}\approx 50$~meV, implying that resonant pumping with the 
${\rm A}$~exciton brings the MoS$_{2}$ monolayer toward a nonequilibrium 
steady-state characterized by a vertical compression 
of the sulphur atoms ($\mathsf{x}_{A_{1}'}<0$). 

As the analytic formula in Eq.~(\ref{exct}) 
has been derived in the limit of low excitation 
density and short pump duration (displacive regime), we have 
 explored the extent to which its applicability
is affected when the photoexcitation deviates from these conditions, 
see Fig.~\ref{fig6}. To facilitate comparisons
a $\t$-slice of Fig.~\ref{fig5}c is shown in panel~a.
In panel~b we consider  longer pump pulses with duration $T_{{\rm 
P}}=50$~fs, while still maintaing a small excitation density $n_{{\rm 
x}}=4\times 10^{11}~{\rm cm}^{-2}$. 
In this case a correction due to the effective density 
$n_{A_{1}'}$ must be 
introduced [see Eqs.(\ref{exct}) and (\ref{ut})] since the 
photoexcitation is no longer displacive. The value $n_{A_{1}'}/n = 
a_{A_{1}'}/|\mathsf{x}_{A_{1}'}|= 0.71$ is extracted 
by taking the ratio between 
the amplitude of the oscillations $a_{A_{1}'}$ of the $A_{1}'$ mode 
as obtained from the numerical simulation and the analytic value 
$|\mathsf{x}_{A_{1}'}|=\frac{x_{0A_{1}' }n|G^{\l}_{A_{1}'}|}{\hbar 
\w_{A_{1}'}}$. 
Remarkably, the validity of the analytic formula extends well 
beyond the initially defined applicability 
region.
In panel~c we consider a very long pump $T_{{\rm P}}=100$~fs 
and a relatively high excitation density $n_{{\rm 
x}}=2.3\times 10^{12}~{\rm cm}^{-2}$.
Applying the correction 
$n_{A_{1}'}/n= 0.10$, the agreement between the simulated 
and predicted excitonic oscillations is still reasonably good.
Further increasing the excitation density, 
$n_{{\rm x}}=5.6\times 10^{12}~{\rm cm}^{-2}$, the 
agreement begins to deteriorate, even within the displacive regime, see 
panel~d, where $n_{A_{1}'}/n = 1$.

\section{Discussion}

We   present the many-body theory of the interaction between 
 resonantly pumped excitons and coherent phonons, and 
derive the 
expression of the X{\it-cph} coupling.
The magnitude of this coupling can exceed that of the X{\it-ph}
coupling in three dimensions, given that the bare (screened) {\it e-ph} 
matrix elements are 
used to calculate the former (latter).
We  also derive a readily applicable formula 
to quantify the modulations of the excitonic energies
in resonantly pumped materials~\cite{trovatello,cohexp2,cohexp3,cohexp4,cohexp5}.
The accuracy of the formula   is demonstrated 
through comparisons with 
model simulations
of transient absorption in a MoS$_{2}$ monolayer.
Our findings provide a straightforward means to directly extract the value of the  
X{\it-cph} coupling, along with accurately determining the magnitude of the excitation density.

\newpage

 \section{Methods}
 {\small
 {\bf Real-time HSEX + Eherenfest method}
 
 We describe in detail the method 
that we use to propagate  in time
the electronic one-particle density matrix
$\r_{\blk ij}$ together  with the phonon displacement $x_{\n}(t)$ and momentum 
$p_{\n}(t)$. Here $\blk$ denotes the cristal momentum,
$i,j$ the spin-band indices that can be valence ($v$) or conduction 
$(c)$, while
$\n$ denotes the phonon mode.
Our method is based on the Hartree plus statically screened 
exchange (HSEX) plus Ehrenfest approximation of many-body theory.
The coupled equations of motion to be propagated in time 
read~\cite{stefanucci_in-and-out_2023,stefanucci_semiconductor_2024}
\bea
i\frac{d}{dt}\rho_{\blk 
ij}(t)&=&[h^{\rm{qp}}_\blk(t),\rho_\blk(t)]_{ij}+I^{\rm{coll}}_{\blk ij}(t),
\nonumber \\
p_{\nu}(t)&=&-\omega_{\nu}x_\nu(t)-\sum_{\blk,ij}g^{\nu}_{ij}(\blk,0)\rho_{\blk ji}(t)  \nonumber \\
\frac{d}{dt} x_{\nu}(t) 
&=&\omega_{\nu}p_{\nu}(t).
\label{eom}
\eea
In Eq.~(\ref{eom}) $h^{\rm qp}$ is the quasi-particle hamiltonian
given by
\bea
h^{\rm{qp}}_{\blk ij}(t)&=& h^{\rm{HSEX}}_{\blk ij}(t) + 
h^{\rm{Eh}}_{\blk ij}(t) \nonumber \\
&=& \d_{ij}\e_{ \blk i}
+\sum_{\blk_{1}mn}
(V_{imnj}^{\bf{0} \blk\blk_{1}}-V_{imjn}^{(\blk-\blk_{1}) \blk 
\blk_{1}})\d\r_{\blk_{1}nm}(t) + \mathbf{E}(t)\cdot \mathbf{d}_{\blk 
ij }+\sum_{\n} x_{\n} (t) g^{\nu}_{ij}(\blk)
\label{hf}
\eea
This hamiltonian accounts for
 (i) the electron kinetic energy (via the band structure $\e_{ \blk i}$), (ii) the interaction with the 
external laser pulses $\mathbf{E}(t)$ (via the  dipole matrix elements 
$\mathbf{d}_{\blk ij }$), (iii) the HSEX self-energy (via the electron-electron interaction $V_{imjn}^{\blq\blk \blk'}$),
and (iv) the Ehrenfest potential describing the coupling of electrons 
with the nuclear displacements
(via the bare electron-phonon coupling $g^{\nu}_{ij}(\blk)\equiv 
g^{\nu}_{ij}(\blk,0)$ at vanishing momentum transfer).
The presence of the HSEX and Ehrenfest potentials
guarantees that strong excitonic effects and the coupling of excitons to coherent 
phonons are included in the photoexcited dynamics.
The SEX interaction is modeled by the 
Rytova-Keldysh potential 
~\cite{keldysh,PhysRevB.84.085406} 
(see also below) which has demonstrated remarkable accuracy in reproducing excitonic binding 
energies in 2D materials~\cite{PhysRevB.88.045318,Chernikov2014,Steinhoff2014}.
The evaluation of the Ehrenfest potential and the effective 
forces acting on the nuclei  
require only the intra-band electron-phonon couplings $g^{\nu}_{ii}(\blk)$ at $\blq=0$~\cite{stefanucci_in-and-out_2023}.
The  couplings
$g$ can be evaluated according to
\be
g^{\nu}_{ii}(\blk)=\sqrt{\frac{\hbar}{2 M \omega_{\nu}}} 
\frac{\partial E_{\blk i}}{\partial x_{\nu}},
\label{gdress}
\ee
where $E_{\blk i}$ is the Kohn-Sham energy of the $i$-th band at momentum $\blk$,
$\omega_{\nu}=\omega_{\nu}(\blq =0)$  is the frequency of mode $\nu$
at the $\Gamma$ point, and $x_{\nu}$ denotes the ion displacement 
along the phonon mode $\nu$ with momentum $\blq=0$.

In MoS$_{2}$ monolayers the mostly coupled phonon is the out-of-plane 
$A'_{1}$ optical mode~\cite{trovatello}.  Therefore, 
only this specific normal mode has been included in our simulations.
The last term in the equation of motion for $\r$ is the collision 
integral $I^{\rm{coll}}(t)$, that incorporates
dynamical correlation effects responsible for, e.g., inelastic 
quasi-particle scattering,
decoherence and dynamical 
screening~\cite{PhysRevLett.127.036402,PhysRevLett.128.016801,PhysRevB.105.125134,perfetto_2023}.
At low and moderate excitation densities the dominant contribution to the collision integral
stems from electron-phonon scattering
which  drives the system toward a depolarized 
regime characterized by the  disappearance of valence-conduction elements
$\rho_{\blk, vc}$ of the density 
matrix~\cite{PhysRevB.103.245103,stefanucci_semiconductor_2024,perfetto_2023}.
For this reason all matrix elements 
of the collision integral are neglected, except for the valence-conduction 
ones that we approximate  as $I^{\rm{coll}}_{\blk, vc}(t)=-2i 
\gamma \rho_{\blk,cv}(t)$.
Here $\hbar/\gamma=250$~fs is the dephasing time of the 
electronic polarization experimentally observed in MoS$_{2}$ at low 
temperature~\cite{dephasing1}.
The three coupled equations of motion in Eq.~(\ref{eom})
are numerically integrated by using a 4th order Runge-Kutta solver with 
a time step $\Delta t = 0.1$~fs as implemented in the CHEERS 
code~\cite{PS-cheers}.

For our simulations in MoS$_{2}$  
we select as active space the two highest valence and the 
two lowest conduction spin-bands. Therefore the spin-band index $i$ can take 
the 4 values $i=\{(v\ua),\,(v\da),\,(c\ua),\,(c\da)\}$.
In this way spin-orbit effects responsible
for the energy splitting between A and B excitons are included. 

From a knowledge of the electronic density matrix we can evaluate the momentum 
resolved carrier populations $n_{\blk i}(t)$ according to
\be
n_{\blk i}(t) =\r_{\blk ii}(t).
\ee
Using the off-diagonal elements of $\r$ we can also
calculate the probe-induced dipole moment
according to
\be
\bld(t)=\sum_{\blk}\bld_{\blk}(t)  , \quad  \bld_{\blk}(t) 
=\sum_{ij} \r_{\blk ij}(t)  \bld_{\blk ji}
\ee

{\bf Modelization of the MoS$_{2}$ monolayer}

We calculate the spin-orbit dependent band structure of a MoS$_{2}$ monolayer 
from the tight-binding parametrization provided in 
Ref.~\cite{PhysRevB.88.085433} that gives low-energy bands $\e_{\blk i}$ 
in very good agreement 
with DFT calculations.
The conduction bands are rigidly shifted-up by 0.6~eV, to match the 
quasiparticle bandgap measured in ARPES experiments~\cite{bandgap1}. 
For the RT simulations we consider the 
two highest valence bands and the two lowest conduction bands as our 
active space.
The Coulomb integral $V_{imjn}^{\blq\blk 
\blk'}$ accounting for the scattering amplitude of two electrons in bands $j$ and $n$ with quasimomenta
$\blk'+\blq$ and $\blk-\blq$ to end up in bands $m$ and $i$ with 
quasimomenta $\blk'$
and $\blk$ respectively  is calculated according to  $V_{imjn}^{\blq\blk 
\blk'}=v_{q} (\blU_{ \blk i}^{\dag} \cdot \blU_{ \blk -\blq n})   (\blU_{ 
\blk ' m}^{\dag} \cdot \blU_{ \blk ' +\blq j}) 
$~\cite{PhysRevB.91.075310},
where $\blU_{ \blk i}$ are the eigenvectors of the Bloch Hamiltonian 
$H_{\blk}$   
and $v_{q}$ is the Rytova-Keldysh 
potential~\cite{keldysh,PhysRevB.84.085406} in momentum space, i.e.
\be
v_{q}=\frac{2\pi}{\e q(1+r_{0}q)} \,.
\ee 
In the above equation   $q=|\blq|$, $r_{0}=33.875~\mathring{\rm 
A}/\e$~\cite{PhysRevB.91.075310}, and  $\e=2.5$ is the dielectric constant of a typical 
substrate (the topstrate is supposed to be air).
The regularization of the diverging value of $v_{q}$ at the $\Gamma$ 
point is expressed as~\cite{PhysRevB.88.245309,PhysRevB.97.205409} 
$v(0)\to \frac{1}{\Omega} \int_{\Omega}d\blq v(q)$.
Here
$\Omega$ represents a 2D domain around $\blq=0$ of linear dimension determined by the discretization
step of the first Brillouin zone. In this study, the first Brillouin 
zone has been discretized using a  $C_{6v}$-symmetric 
grid comprising 3072 $\mathbf{k}$-points.
In semiconductor materials the Coulomb integrals that do not conserve the 
number of valence and conduction electrons
are typically negligible~\cite{Groenewald2016} and have been  
excluded from our calculations. 
Finally the  
dipole  matrix elements are evaluated according 
to~\cite{PhysRevB.80.085117,Steinhoff2014}
\be
\bld_{ \blk ij}=\frac{1}{i}\frac{1}{\e_{\blk i}-\e_{\blk j}} \blU_{ 
\blk i}^{\dagger}\cdot \partial_{\blk}H_{\blk} \cdot  \blU_{ \blk j}. 
\ee

Concerning the phonon input, the electron-phonon coupling in Eq.~(\ref{gdress}) 
are obtained by displacing manually the ions along the phonon eigenvector directions
(with equilibrium lattice constant $a=3.12~\AA$) and by calculating 
the corresponding difference in the band structure.
This procedure is very accurate for the out-of-plane optical mode 
$A_{1}'$~\cite{ehrmos2}.
The Kohn-Sham energies $E_{\blk i}$ for a given displaced configuration are
calculated with Quantum ESPRESSO using LDA in a $24\times 24$ grid. The 
phonon frequencies $\w_{\n}$ are evaluated within the same grid.

 }

\newpage

\vspace{1cm}

{\bf Data availability}

The datasets used for this study are available from the corresponding 
authors upon request.

{\bf Code availability}

The code used for this study is available from the corresponding 
authors upon request.

{\bf Acknowledgments}

The Authors acknowledge funding from Minestero Universit\`a e Ricerca 
PRIN under grant agreement
No. 2022WZ8LME, from the INFN TIME2QUEST project, 
from the European Research Council  
MSCA-ITN TIMES under grant agreement 101118915, and 
from Tor Vergata University through Project TESLA.

{\bf Author contributions}

KW performed the DFT ab-initio calculations of the electron-phonon 
couplings. EP and GS contributed equally to this work and to the preparation of this manuscript.

{\bf Competing interests}

The Authors declare no competing interests.

\newpage

\vspace{1cm}
\begin{center}
\large
{\bf References}
\end{center}

\bibliographystyle{unsrtnat}

\end{document}